# Robustness of networks with topologies of dependency links

Yuansheng Lin[1,2], Daqing Li[1,2,*], Rui Kang[1,2], Shlomo Havlin[3]

[1] School of Reliability and Systems Engineering, Beihang University, Beijing 100191, China

[2] Science and Technology on Reliability and Environmental Engineering Laboratory, Beijing 100191, China

[3] Department of Physics, Bar-Ilan University, Ramat Gan 5290002, Israel

* li.daqing.biu@gmail.com

**Abstract**

The robustness of complex networks with dependencies has been studied in recent years. However, previous studies focused on the robustness of networks composed of dependency links without network topology. In this study, we will analyze the percolation properties of a realistic network model where dependency links follow certain network topology. We perform the theoretical analysis and numerical simulations to show the critical effects of topology of dependency links on robustness of complex networks. For Erdös-Rényi (ER) connectivity network, we find that the system with dependency of RR topology is more vulnerable than system with dependency of ER topology. And RR-RR (i.e. random-regular (RR) network with dependency of RR topology) disintegrates in an abrupt transition. In particular, we find that the system of RR-ER shows different types of phase transitions. For system of different combinations, the type of percolation depends on the interaction between connectivity topology and dependency topology.

## 1. Introduction

Components in critical infrastructures cannot function independently, which interact with others through different dependencies [1-14]. For example, nodes in power grid convey power flow to certain loads, whose faults will induce the overload to other nodes. This overload can be considered as certain dependency between nodes, whose failure will cause another node to failure. Initiated by random failures, these dependencies may lead to catastrophic events including blackouts in power grid and jamming in transportation networks [15].

In the previous studies, dependency links could be mainly classified into two categories: (1) serial: when a node fails, it will induce the failures of other members having dependency link with it. Parshani et al. [8] first introduce a network model having dependency groups with fixed size 2. Bashan et al. [5] generalize Parshani's results and also study the effects of dependency groups whose sizes are characterized by different classic distributions, such as normal distribution or Poisson distribution, in a single network; (2) parallel: as described in Ref. [9], a node could have multiple support dependency relations from other network, where the failure of this node will be caused only by the failures of all its dependency neighbors.

However, in realistic cascading failures [14], dependencies between nodes in real systems can be complicated and form certain topology, which is not considered in the existing study on networks with dependency. Different from simple dependency structures (Fig.1 (a)), here we study the dependency topology (Fig.1 (b)) by analyzing the percolation properties of networks with different topologies of dependency links. We present a general analytical framework to analyze the critical effects of topology of dependency links on robustness of complex networks. And the corresponding theoretical results are confirmed by simulations. We take ER network and RR network as the reference cases, and find that ER connectivity network with dependency of RR topology is more fragile than system with dependency of ER topology. And RR-RR collapses in a form of first order with abrupt change. For RR-ER, the type of phase transition is relying on the average connectivity degree and the density of dependency links.

This paper is organized as follows. In the second section, we will introduce the network model composed of dependency links following certain topology. In the third section, we present both the simulation and analytical results of cascading failures in ER network and RR network with different dependency topologies. In section 4, we summarize our results.

## 2. Model description

We study cascading failures in networks containing different topologies of dependency links. The iterative process between percolation stage and dependency stage are as follows [5, 16]:

1. Initially, we randomly remove a fraction $1-p$ of nodes from the connectivity network;

2. Percolation stage: in connectivity network, nodes out of the giant component will be removed;

3. Dependency stage: If there is one node failed, all the other nodes in the same dependency cluster will become failed.

4. A nodes fails at the percolation stage will lead to the failure of the entire dependency cluster, which in turn motivates the next percolation stage. Then a cascade of failures will continue until no additional failure occurs.

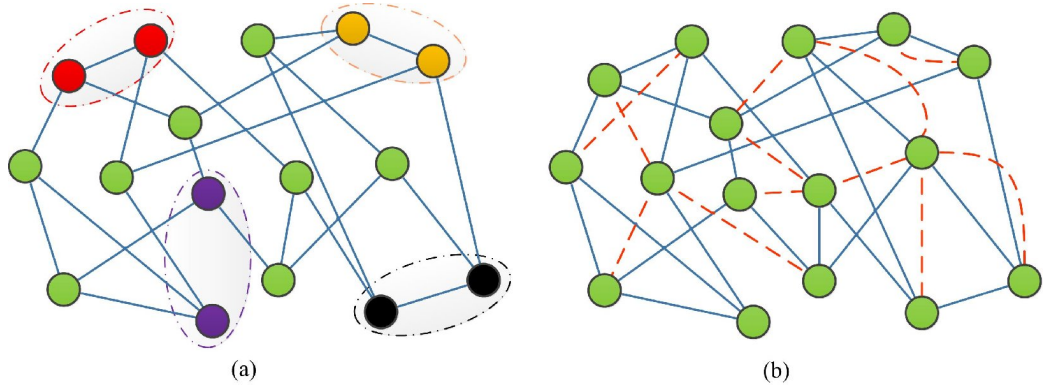

Figure 1 Network with different dependency structures. (a) Simple dependency structure: connectivity network with dependency groups of size 2: The solid blue lines represent connectivity links, and dependency groups are encompassed by the dash dot line which has same color with nodes in the corresponding group; (b) topological dependency structure: connectivity network with topology of dependency links, where the solid blue lines indicate connectivity links and red dash lines represent topology of dependency links.

## 3. Results

We take ER network and RR network as the reference cases, which could be solved explicitly, and discuss the robustness of network with different topologies of dependency links:

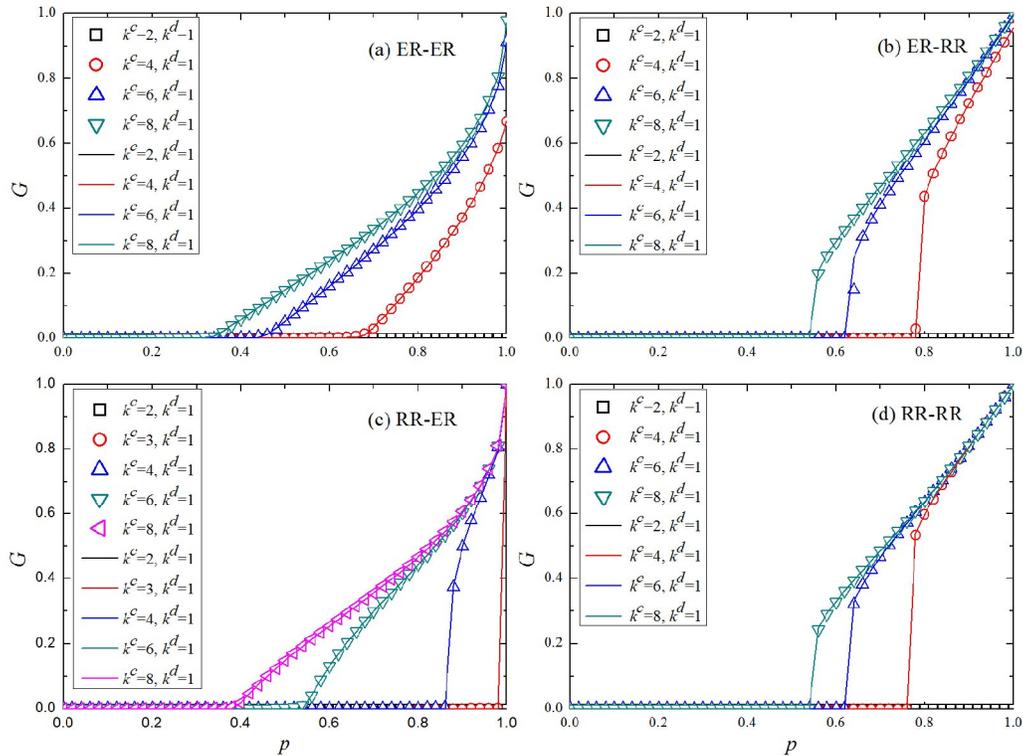

Figure 2 The size of largest component, $G$, vs $p$, the remaining fraction of functional nodes after initial

random attack. For different cases: (a) ER-ER; (b) ER-RR; (c)RR-ER; (d)RR-RR, all open symbols represent the numerical results of systems of 100000 nodes for 200 realizations. The solid curves show the theoretical results.

**A. ER network with dependency of ER topology (ER-ER)**

In this case, both connectivity links and dependency links follow ER topology of their own. In ER connectivity network, nodes are connected randomly via connectivity links, and their connectivity degrees are Poisson distributed [17-19]. When their average connectivity degree is $k^c$, we can get $G_0(\xi) = G_1(\xi) = \exp[k^c(\xi-1)]$. Analogously, we suppose that the average dependency degree of dependency network is $k^d$. Then $G_0(\eta) = G_1(\eta) = \exp[k^d(\eta-1)]$. We can gain the theoretical results of the size of giant component $G$ after cascading failure triggered by initially removed $1-p$ fraction of nodes. Our simulation results of ER-ER agree with the theoretical predictions (as illustrated in Fig.2 (a)). We find that the ER-ER system breaks down in a continuous percolation process. In Fig. 2a, we can also find that network resilience is increased with increasing connectivity degree at a constant average dependency degree $k^d$.

**B. ER network with dependency of RR topology (ER-RR)**

Dependency links could also form a RR topology. When the dependency degree of each node is $k^d$, the generating function of the dependency degree distribution is $G_0(\eta) = \eta^{k^d}$ and $G_1(\eta) = \eta^{k^d-1}$. Specially, when $k^d$ is 1, $q(2)=1$. Compared with ER-ER, ER-RR is more vulnerable since the cascading process of ER-RR undergoes an abrupt collapse near threshold (Fig.2 (b)).

**C. RR network with dependency of ER topology (RR-ER)**

In RR-ER, every node has $k^c$ connectivity links with others, thus $G_0(\xi) = \xi^{k^c}$ and $G_1(\xi) = \xi^{k^c-1}$. And for ER topology of dependency links, when its average dependency degree is $k^d$, $G_0(\eta) = G_1(\eta) = \exp[k^d(\eta-1)]$. In Fig.2 (c), we show the analytical and numerical results for RR-ER. Interestingly, the percolation process of RR-ER seems to be characterized by two distinct types of transitions. For low density of connectivity links, e.g. $k^c=4$, the system undergoes an abrupt phase transition. As $k^c$ increases, the system is becoming more stable with a continuous phase transition. This is in contrast to the case of ER-ER where a continuous phase transition characterizes the cascade process of ER-ER.

**D. RR network with dependency of RR topology (RR-RR)**

Nodes in RR connectivity network are also connected by dependency links following RR topology. So $G_0(\xi) = \xi^{k^c}$ and $G_1(\xi) = \xi^{k^c-1}$. At the same time, for dependency links of RR topology, $G_0(\eta) = \eta^{k^d}$ and $G_1(\eta) = \eta^{k^d-1}$. In Fig.2 (d), RR-RR disintegrates in an abrupt transition. For $k^d = 1$, when $k^c$ is slightly larger, such as $k^c = 4$ or $k^c = 6$, RR-RR is more vulnerable than RR-ER as RR-ER breaks down in a continuous transition.

As found in these different combinations, the transition type of percolation is determined by the interaction between connectivity topology and dependency topology.

Finally, we show how the average connectivity degree $k^c$ and the average dependency degree $k^d$ affect the robustness of RR-ER by measuring the critical points $p^c$ of phase transition (Fig.3). For different types of phase transitions, the numerical results of $p_c$ are obtained by different means. In RR-ER, for a constant average dependency degree $k^d$, when $k^c$ is small, the system disintegrates in a form of first order, and we identify the critical points when the number of iterative (NOI) failures reaches maximum [8]. When $k^c$ gets larger, the system become more robust, and it leads to a second order phase transitions. And the critical point for second-order transition is identified when the size of the second largest component approaches the maximum value. Fig.3 also show that more dependency links will make the cascading process of the system change from the second order into the first order, e.g. when $k^c$ is 4. And for a large $k^c$, e.g. $k^c$ =8, though the system breaks down in a continuous way, higher density of dependency links will make system easier to fragment.

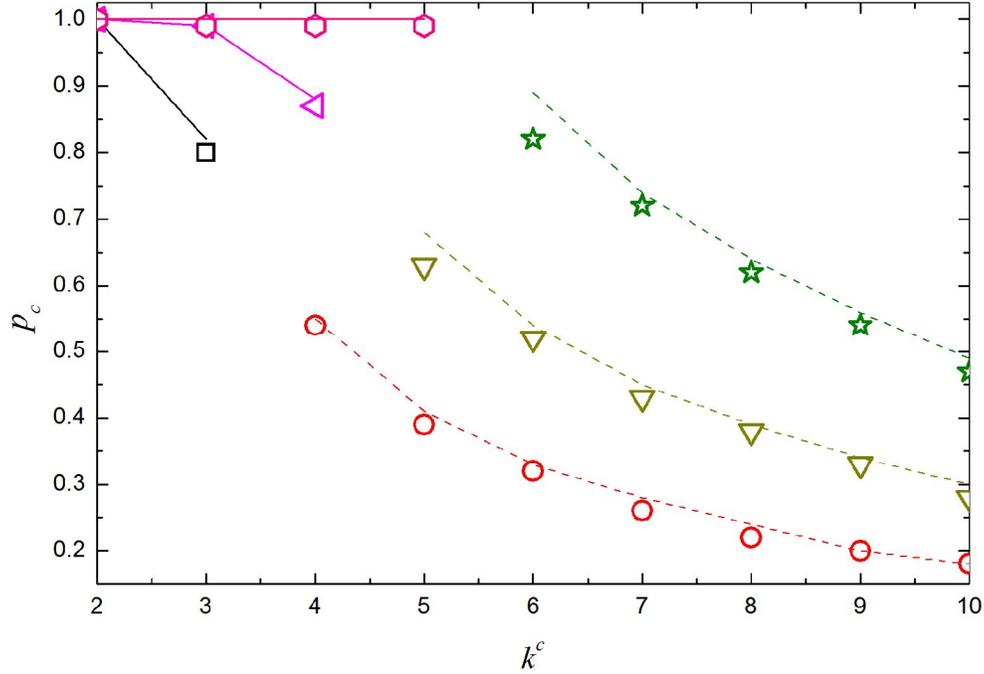

Figure 3 Theoretical and simulation value of the remaining critical fraction of functional nodes $p_c$ are compared for RR-ER with different average connectivity degrees ($k^c$) and different average dependency degrees ($k^d$). The solid lines show the theoretical results for the first order cases, and the dash lines are the cases of the second order. Open symbols are gained by simulation techniques. The corresponding $k^d$ from left to right is 0.5, 1 and 1.5.

## 4. Conclusions

Here we study the percolation properties of networks with dependency links following certain topology, rather than simple form of dependency structures. We propose the theoretical and numerical analysis of percolation properties including size of giant component, order of phase transition and critical threshold. In particular, we study the robustness of ER network and RR network with different topologies of dependency links and predict their critical thresholds of phase transitions. In ER connectivity network, we find that the system with dependency links of RR topology will be more vulnerable than system with dependency links of ER topology. And for RR connectivity network with dependency of RR topology, the system disintegrates in an abrupt transition. Especially, we find that the system of RR-ER shows different types of phase transition, which depends on its connectivity degree and dependency degree. And we identify the critical thresholds and distinguish the types of phase transitions using the graphical solution method. Although our theoretical results are illustrated for ER and RR networks, they can be applied for other random networks.